# O(1) Delta Component Computation Technique for the Quadratic Assignment Problem

Sergey Podolsky, Yuri Zorin
National Technical University of Ukraine “Kyiv Polytechnic Institute”
Faculty of Applied Mathematics
Computer Systems Software Department
sergey.podolsky@gmail.com

## 1. Abstract

The paper describes a novel technique that allows to reduce by half the number of delta values that were required to be computed with complexity O(N) in most of the heuristics for the quadratic assignment problem. Using the correlation between the old and new delta values, obtained in this work, a new formula of complexity O(1) is proposed. Found result leads up to 25% performance increase in such well-known algorithms as Robust Tabu Search and others based on it.

## 2. Introduction

The quadratic assignment problem (QAP) was first mentioned by Koopmans and Beckmann in 1957 [1] and still remains one of the most challenging combinatorial optimization problems. The problem formulation is as follows. There exist $N$ locations and $N$ facilities. Distances between each pair of the locations and flows (i.e. number of transportations) between each pair of facilities are provided. The goal is to locate (assign) all the facilities into different locations such as to minimize the sum of all distances multiplied by the corresponding flows. Mathematically, this could be formulated as an objective function:

$$\min \sum_{i=1}^{N} \sum_{j=1}^{N} d_{ij} f_{\pi_i \pi_j}, \tag{1}$$

where $\pi_x$ is a facility assigned to location $x$; $d_{ij}$ is a distance between locations $i$ and $j$; $f_{\pi_i \pi_j}$ is a flow between facilities $\pi_i$ and $\pi_j$. Since the problem is NP-hard [2] there is no exact algorithm that could solve the QAP in polynomial time. Furthermore, the travelling salesman problem (TSP) may be considered as a special case of the QAP for the case where all of the facilities are connected via flows of constant value into a single ring, while other flows are set to zero. Thus, QAP could be considered more challenging than TSP.

Only heuristic methods allow to obtain feasible solutions for QAP instances of size 30 and higher in reasonable time. One of the most efficient heuristic algorithms for the QAP is the Robust Tabu Search (Ro-TS) by Eric Taillard [3], which produces high quality solutions even for very large instances. Among other successful heuristics are genetic algorithms [4], ant systems [5] and others. Typically, the predominant majority of them are based on a similar approach of the neighborhood representation obtained by pairwise exchange of elements in the solution vector.

## 3. Background of the Neighborhood Scanning

The process of neighborhood scanning defined in Ro-TS is one of the most representative environments subject to improvement, since Ro-TS performs initial solution construction and objective function (1) evaluation only once on the beginning of the algorithm before main iterations start. During the initial stage, a random solution vector is generated and its cost is computed by objective function as defined by (1). On each main iteration of the Ro-TS, a move, which relies on the exchange of two elements from the current solution vector is performed. A pair of elements to be exchanged is selected among all $C_N^2$ pairs in accordance with $\Delta_{ij}$ minimization criterion. The actual selection criterion is more complicated than just a delta value minimization, its discussion is omitted. After exchange, a new cost could be obtained from the previous cost by addition of a $\Delta_{ij}$ term, where all $C_N^2$ values $\Delta_{ij}$ are computed using the next formula during the initial stage of the Ro-TS (each with time complexity O(N)) before the main iterations:

$$\Delta_{ij} := (d_{ii} - d_{jj})\left(f_{\pi_j\pi_j} - f_{\pi_i\pi_i}\right) + (d_{ij} - d_{ji})\left(f_{\pi_j\pi_i} - f_{\pi_i\pi_j}\right) +$$
$$+ \sum_{\substack{g=1,\\ g\neq i,j}}^{N} \left[(d_{gi} - d_{gj})\left(f_{\pi_g\pi_j} - f_{\pi_g\pi_i}\right) + (d_{ig} - d_{jg})\left(f_{\pi_j\pi_g} - f_{\pi_i\pi_g}\right)\right]. \quad (2)$$

It is convenient to store all $C_N^2$ computed values $\Delta_{ij}$ in a jagged matrix and fetch them to modify solution cost value after elements $i$ and $j$ exchange. Each time after the two elements exchange, all $C_N^2$ values $\Delta_{ij}$ become unfeasible and need to be corrected. Clearly, the new $\Delta_{pq}$ value (after the elements $p$ and $q$ just have been exchanged) is evaluated trivially as $-\Delta_{pq}$, reflecting the reverse exchange of the same pair. For all $\Delta_{ij}$ which do not involve the two elements $p$ and $q$ that were exchanged, i.e. $\{i,j\} \cap \{p,q\} = \{\emptyset\}$, new values $\Delta_{ij}^*$ can be evaluated using their old values $\Delta_{ij}$ with complexity O(1):

$$\Delta_{ij}^* := \Delta_{ij} + (d_{pi} - d_{pj} + d_{qj} - d_{qi})\left(f_{\pi_p\pi_j} - f_{\pi_p\pi_i} + f_{\pi_q\pi_i} - f_{\pi_q\pi_j}\right) +$$
$$+(d_{ip} - d_{jp} + d_{jq} - d_{iq})\left(f_{\pi_j\pi_p} - f_{\pi_i\pi_p} + f_{\pi_i\pi_q} - f_{\pi_j\pi_q}\right).$$

However, as has been observed by Taillard [3], for those $\Delta_{ij}$ which involve one of the indices $p$ or $q$ from the last exchange, i.e. $|\{i,j\} \cap \{p,q\}| = 1$, all new values $\Delta_{ij}^*$ must be evaluated anew with complexity O(N) by (2). This technique of complete delta recomputation became widespread since it was originally described by Taillard and proceeded without much improvement into further research such as that of Reactive Tabu Search [6] and is still used nowadays in miscellaneous Ro-TS implementations such as for sparse matrices [7] etc. This paper reveals how to bypass this issue and to evaluate with linear time complexity only half of deltas involving nodes from the last pair exchange.

## 4. Significance and Performance

As mentioned before, the total count of all possible exchanges equals $C_N^2 = \frac{N(N-1)}{2}$. Among them there are $2(N-2)$ possible pair exchanges that involve one of the two just exchanged elements. The last count is obtained from considerations that each of the two elements that were swapped can be exchanged further with each element from the rest of $N-2$ other distinct elements, as depicted on Figure 1.

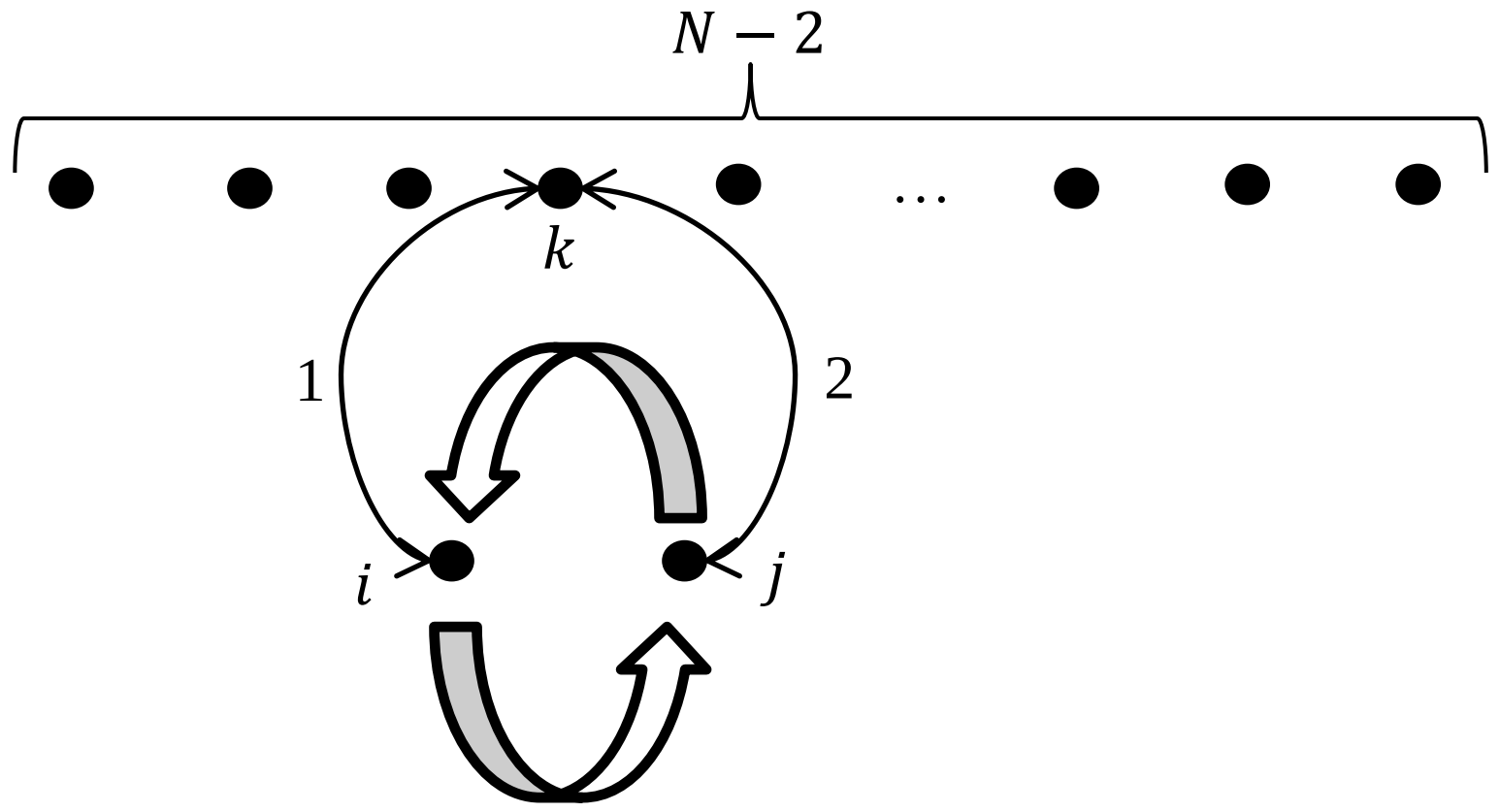

**Figure 1. An illustration of further exchange of one of the 2 swapped elements**

Consequently, the total count of those $\Delta_{ij}$ that could be corrected with complexity O(1) is equal to the difference of all possible exchanges count and the total count of exchanges being corrected with complexity O(N). This difference is expressed as

$$\frac{N(N-1)}{2} - 2(N-2) = \frac{(N-4)(N-1)}{2} + 2.$$

Thus, we are required to compute only $N-2$ instead of $2(N-2)$ new delta values with complexity O(N) and the rest with O(1). However, it is still unclear if this will significantly affect the iteration performance. The answer was obtained after the Ro-TS open source C++ code was profiled using MS Visual Studio Profiling tools on `Tai100a` QAP instance. As shown on Figure 2, the `compute_delta` method which computes deltas anew takes over 50% of total samples being the most expensive call, while `compute_delta_part` takes only 29% of total samples to update each of other deltas with O(1) time complexity. Therefore, we suggest that computational time per iteration could be successfully reduced by up to a quarter using the proposed technique, especially on large QAP instances.

**Hot Path**

The most expensive call path based on sample counts

| Function Name | Inclusive Samples % | Exclusive Samples % |
|---|---|---|
| QuadraticAssignmentProblem.exe | 100,00 | 0,00 |
| __tmainCRTStartup | 100,00 | 0,00 |
| _wmain | 100,00 | 0,00 |
| TaillardTabuQap::solve(int,long,long) | 99,87 | 18,04 |
| **TaillardTabuQap::compute_delta(int,int * *,int * *,int *,int,int)** | **52,73** | **52,73** |

Related Views: Call Tree Functions

**Functions Doing Most Individual Work**

Functions with the most exclusive samples taken

| Name | Exclusive Samples % |
|---|---|
| TaillardTabuQap::compute_delta(int,int * *,int * *,int *,int,int) | 52,73 |
| TaillardTabuQap::compute_delta_part(int * *,int * *,int *,int * *,int,int,int,int) | 29,07 |
| TaillardTabuQap::solve(int,long,long) | 18,04 |
| std::basic_istream<char,struct std::char_traits<char> >::operator>>(int &) | 0,13 |
| TaillardTabuQap::rando(void) | 0,03 |

**Figure 2. Sample profiling report for Ro-TS on Tai100a instance**

## 5. Novel O(1) Delta Component Computation Technique Exploration

Suppose we have three elements with indices $i$, $j$, $k$ in the solution vector and the values $\Delta_{ij}$, $\Delta_{ik}$, $\Delta_{jk}$ are known. Our purpose is to exchange a pair of elements $i$ and $j$ and to compute new values $\Delta^*_{ik}$ and $\Delta^*_{jk}$ afterwards. First, let's consider the assignments[1] of flows $f_{\pi_i\pi_g}$, $f_{\pi_j\pi_g}$, $f_{\pi_k\pi_g}$ to distances $d_{ig}$, $d_{jg}$, $d_{kg}$, where by $g$ we denote an arbitrary element from the rest of $N-3$ elements distinct from $i$, $j$, $k$ (see Figure 3). All the values $\Delta_{ij}$, $\Delta_{ik}$, $\Delta_{jk}$ include the following terms that indicate the cost change caused by reassignments of flows to distances to (from) element $g$ after corresponding pairs exchange:

$$\delta_{ij} = -d_{ig}f_{\pi_i\pi_g} - d_{jg}f_{\pi_j\pi_g} + d_{ig}f_{\pi_j\pi_g} + d_{jg}f_{\pi_i\pi_g},$$

$$\delta_{ik} = -d_{ig}f_{\pi_i\pi_g} - d_{kg}f_{\pi_k\pi_g} + d_{ig}f_{\pi_k\pi_g} + d_{kg}f_{\pi_i\pi_g},$$

$$\delta_{jk} = -d_{jg}f_{\pi_j\pi_g} - d_{kg}f_{\pi_k\pi_g} + d_{jg}f_{\pi_k\pi_g} + d_{kg}f_{\pi_j\pi_g}.$$

Let's consider the assignments of the same three flows to the same three distances after the elements $i$ and $j$ are exchanged (see Figure 4). The new values $\Delta^*_{ij}$, $\Delta^*_{ik}$, $\Delta^*_{jk}$ that we need to compute after this exchange will contain the following terms:

$$\delta^*_{ij} = -\delta_{ij},$$

$$\delta^*_{ik} = -d_{ig}f_{\pi_j\pi_g} - d_{kg}f_{\pi_k\pi_g} + d_{ig}f_{\pi_k\pi_g} + d_{kg}f_{\pi_j\pi_g},$$

$$\delta^*_{jk} = -d_{jg}f_{\pi_i\pi_g} - d_{kg}f_{\pi_k\pi_g} + d_{jg}f_{\pi_k\pi_g} + d_{kg}f_{\pi_i\pi_g}.$$

It is important to note, that exactly these $N-3$ terms $\delta^*_{ik}$ and $\delta^*_{jk}$ for each arbitrary element $g$ need computations of complexity O(N) for each new $\Delta^*_{ik}$ and $\Delta^*_{jk}$ values evaluation.

[1] Though in terms of the QAP an assignment usually means the assignment of facility to specific location, here we mention the assignment of flow to distance caused by a pair of regular QAP assignments.

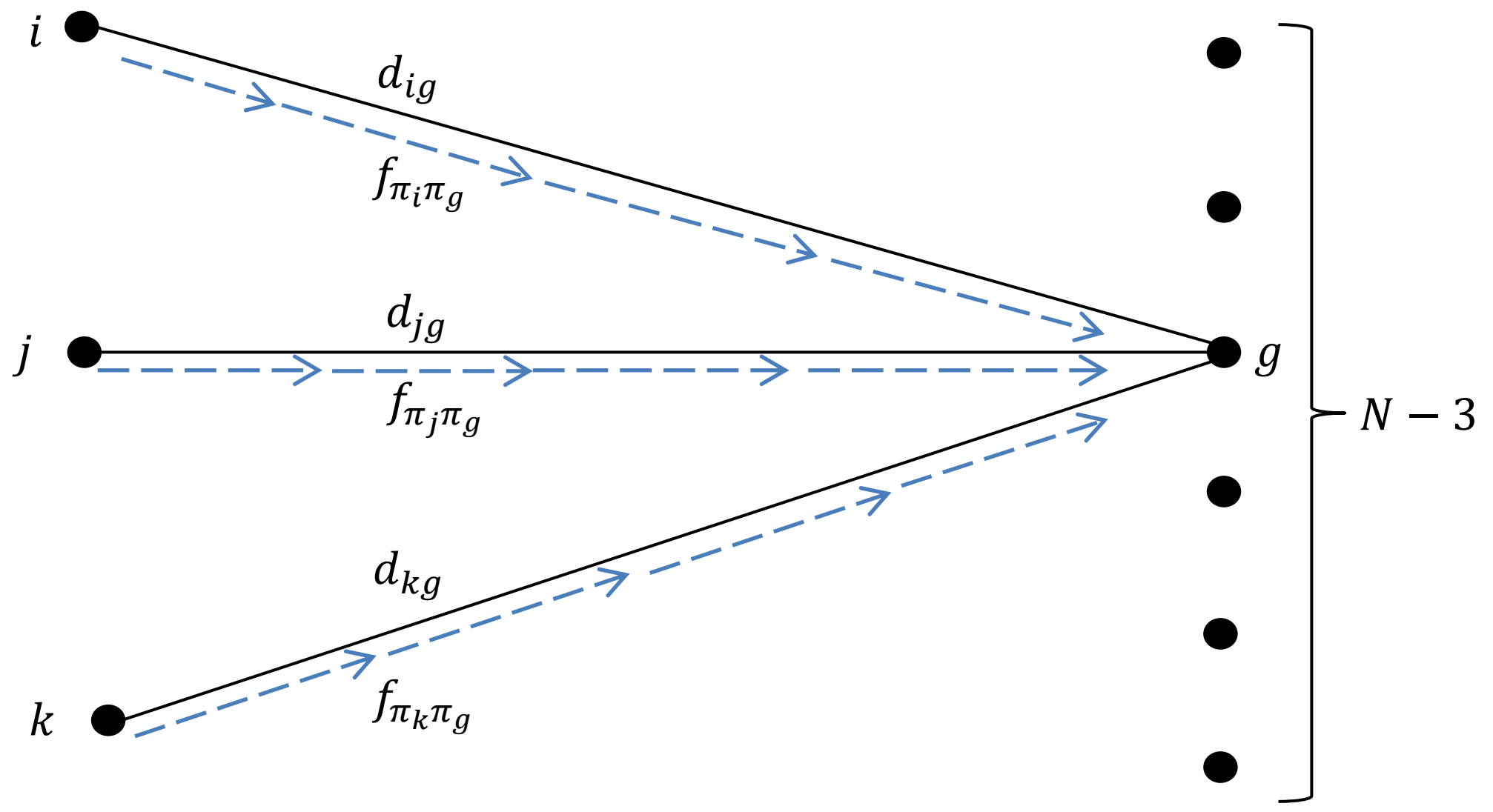

**Figure 3. Assignments of flows to distances before any exchanges**

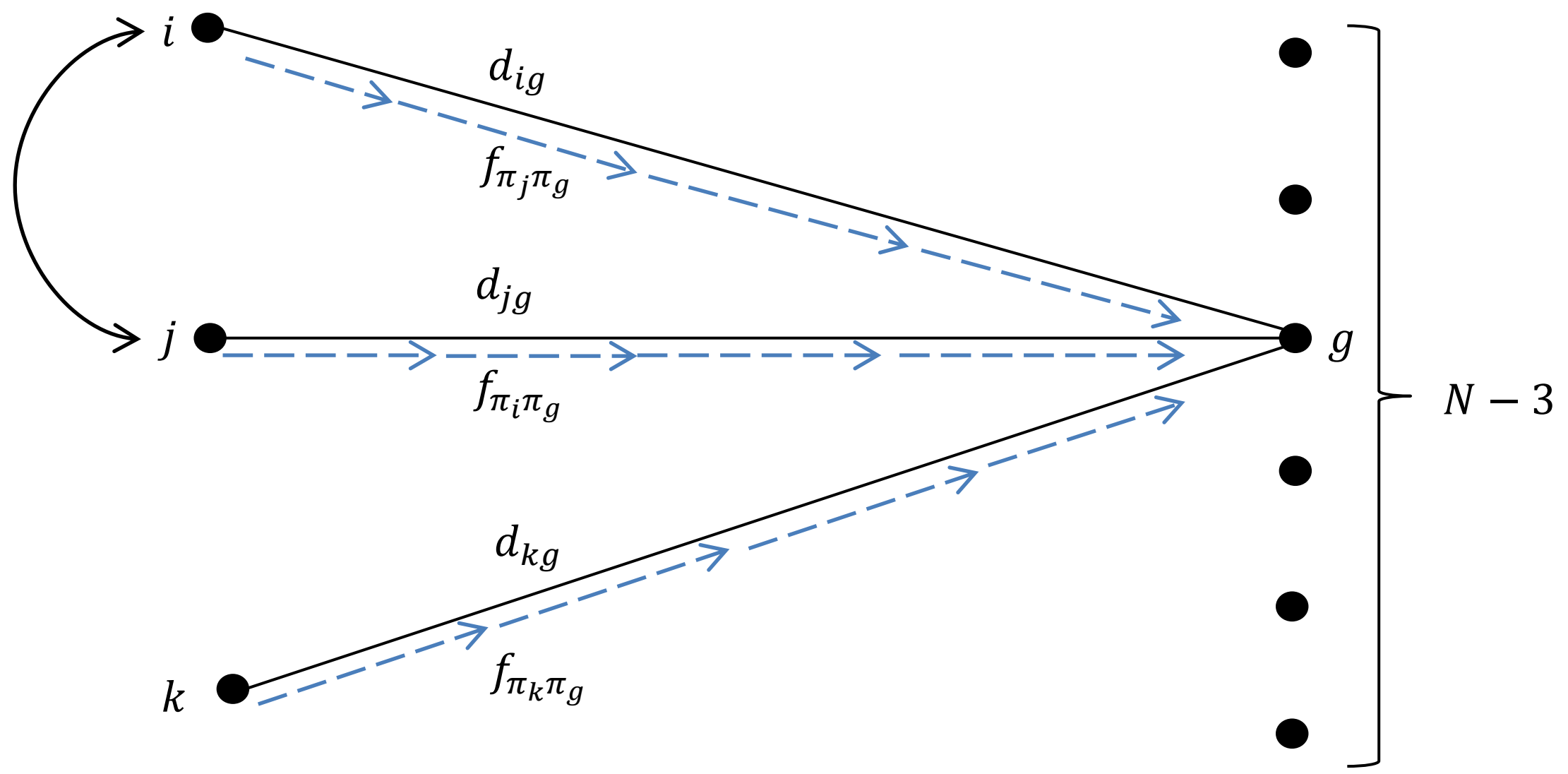

**Figure 4. Assignments of flows to distances after *i* and *j* exchange**

It is trivial to show from the RHS of expressions for $\delta_{ij}$, $\delta_{ik}$, $\delta_{jk}$, $\delta^*_{ik}$, $\delta^*_{jk}$ a dependency between them is formulated as

$$\delta^*_{ik} + \delta^*_{jk} = \delta_{ik} + \delta_{jk} - \delta_{ij}. \quad (3)$$

This result signifies that it is not necessary to compute both new values $\delta^*_{ik}$ and $\delta^*_{jk}$ simultaneously. It is enough to compute anew just one of them and the second one could be evaluated via the first one using previous delta values.

So far we have considered only those terms $\delta_{ij}$, $\delta_{ik}$, $\delta_{jk}$, $\delta^*_{ik}$, $\delta^*_{jk}$ which correspond to connections of our elements $i$, $j$, $k$ with each of other $N-3$ elements. Now, let's consider the terms $R_{ij}$, $R_{ik}$, $R_{jk}$, $R^*_{ik}$, $R^*_{jk}$ which indicate the cost change caused by assignments of flows to distances among the elements $i$, $j$, $k$. The variables $R_{ij}$, $R_{ik}$, $R_{jk}$ indicate the cost change before elements $i$ and $j$ are swapped, while the variables $R^*_{ik}$, $R^*_{jk}$ indicate the cost change after the $i$ and $j$ exchange:

$$\Delta_{ij} = \delta_{ij} + R_{ij},$$

$$\Delta_{ik} = \delta_{ij} + R_{ik},$$

$$\Delta_{jk} = \delta_{jk} + R_{jk},$$

$$\Delta^*_{ik} = \delta^*_{ik} + R^*_{ik},$$

$$\Delta^*_{jk} = \delta^*_{jk} + R^*_{jk}.$$

The values $R_{ij}$, $R_{ik}$, $R_{jk}$, $R^*_{ik}$, $R^*_{jk}$ are expressed by the following formulae:

$$\begin{aligned} R_{ij} &= (d_{ik} - d_{jk})\left(f_{\pi_i\pi_k} - f_{\pi_j\pi_k}\right) + (d_{ki} - d_{kj})\left(f_{\pi_k\pi_i} - f_{\pi_k\pi_j}\right) \\ &+ (d_{ii} - d_{jj})\left(f_{\pi_i\pi_i} - f_{\pi_j\pi_j}\right) + (d_{ij} - d_{ji})\left(f_{\pi_i\pi_j} - f_{\pi_j\pi_i}\right) \end{aligned} \tag{4}$$

$$\begin{aligned} R_{ik} &= (d_{ij} - d_{kj})\left(f_{\pi_k\pi_i} - f_{\pi_j\pi_i}\right) + (d_{ji} - d_{jk})\left(f_{\pi_i\pi_k} - f_{\pi_i\pi_j}\right) \\ &+ (d_{ii} - d_{kk})\left(f_{\pi_k\pi_k} - f_{\pi_j\pi_j}\right) + (d_{ki} - d_{ik})\left(f_{\pi_j\pi_k} - f_{\pi_k\pi_j}\right) \end{aligned} \tag{5}$$

$$\begin{aligned} R_{jk} &= (d_{ki} - d_{ji})\left(f_{\pi_i\pi_j} - f_{\pi_k\pi_j}\right) + (d_{ik} - d_{ij})\left(f_{\pi_j\pi_i} - f_{\pi_j\pi_k}\right) \\ &+ (d_{jj} - d_{kk})(f_{\pi_k\pi_k} - f_{\pi_i\pi_i}) + (d_{kj} - d_{jk})(f_{\pi_i\pi_k} - f_{\pi_k\pi_i}) \end{aligned} \tag{6}$$

$$\begin{aligned} R^*_{ik} &= (d_{ij} - d_{kj})\left(f_{\pi_k\pi_j} - f_{\pi_i\pi_j}\right) + (d_{ji} - d_{jk})\left(f_{\pi_j\pi_k} - f_{\pi_j\pi_i}\right) \\ &+ (d_{ii} - d_{kk})(f_{\pi_k\pi_k} - f_{\pi_i\pi_i}) + (d_{ik} - d_{ki})(f_{\pi_k\pi_i} - f_{\pi_i\pi_k}) \end{aligned} \tag{7}$$

$$\begin{aligned} R^*_{jk} &= (d_{ji} - d_{ki})\left(f_{\pi_k\pi_i} - f_{\pi_j\pi_i}\right) + (d_{ij} - d_{ik})\left(f_{\pi_i\pi_k} - f_{\pi_i\pi_j}\right) \\ &+ (d_{jj} - d_{kk})\left(f_{\pi_k\pi_k} - f_{\pi_j\pi_j}\right) + (d_{jk} - d_{kj})\left(f_{\pi_k\pi_j} - f_{\pi_j\pi_k}\right) \end{aligned} \tag{8}$$

Hence we can represent the equality (3) as

$$(\Delta^*_{ik} - R^*_{ik}) + (\Delta^*_{jk} - R^*_{jk}) = (\Delta_{ik} - R_{ik}) + (\Delta_{jk} - R_{jk}) - (\Delta_{ij} - R_{ij}).$$

Thus, we can compute a new value $\Delta^*_{jk}$ with complexity O(1) using prior computed value $\Delta^*_{ik}$ and old values $\Delta_{ij}$, $\Delta_{ik}$, $\Delta_{jk}$ via formula

$$\Delta^*_{jk} = \Delta_{ik} + \Delta_{jk} - \Delta_{ij} - \Delta^*_{ik} - R_{ik} - R_{jk} + R_{ij} + R^*_{ik} + R^*_{jk}. \tag{9}$$

After the substitution of expressions (4-8) for $R_{ij}$, $R_{ik}$, $R_{jk}$, $R^*_{ik}$, $R^*_{jk}$ into

$$-R_{ik} - R_{jk} + R_{ij} + R^*_{ik} + R^*_{jk},$$

which is a part of (9) RHS without deltas, we will obtain a cumbersome expression (Appendix A) which has no benefits on small QAP instances due to numerous arithmetical operations.

An attempt to simplify such representation was made via MATLAB script utilizing the `simplify` embedded function (Appendix B). As a result the following final reduced formula was obtained to compute all new values $\Delta^*_{jk}$ with time complexity O(1):

$$\begin{aligned} \Delta^*_{jk} &= \Delta_{jk} + \Delta_{ik} - \Delta_{ij} - \Delta^*_{ik} - \\ &-(d_{ij} - d_{ik} - d_{ji} + d_{jk} + d_{ki} - d_{kj})(f_{ij} - f_{ik} - f_{ji} + f_{jk} + f_{ki} - f_{kj}). \end{aligned}$$

## 6. Conclusions

The result obtained in this paper can be successfully implemented into other heuristics that utilize the whole neighbor solutions scanning. Ro-TS is the most representative and exploitable among them, because the solution construction is performed only once and the rest of computational time is used for neighborhood scanning and delta values update. Thus, we can obtain significant performance increase substituting half of the O(N) computation operations with O(1) ones.

# Appendix A. The Final Formula Portion before Simplification

$$
\begin{aligned}
&-R_{ik} - R_{jk} + R_{ij} + R_{ik}^{*} + R_{jk}^{*} = \\
&= \left(d_{ij} - d_{kj}\right)\left(f_{\pi_k\pi_i} - f_{\pi_j\pi_i}\right) + \left(d_{ji} - d_{jk}\right)\left(f_{\pi_i\pi_k} - f_{\pi_i\pi_j}\right) + \\
&+ \left(d_{ii} - d_{kk}\right)\left(f_{\pi_k\pi_k} - f_{\pi_j\pi_j}\right) + \left(d_{ki} - d_{ik}\right)\left(f_{\pi_j\pi_k} - f_{\pi_k\pi_j}\right) - \\
&- \left(d_{ki} - d_{ji}\right)\left(f_{\pi_i\pi_j} - f_{\pi_k\pi_j}\right) + \left(d_{ik} - d_{ij}\right)\left(f_{\pi_j\pi_i} - f_{\pi_j\pi_k}\right) + \\
&+ \left(d_{jj} - d_{kk}\right)\left(f_{\pi_k\pi_k} - f_{\pi_i\pi_i}\right) + \left(d_{kj} - d_{jk}\right)\left(f_{\pi_i\pi_k} - f_{\pi_k\pi_i}\right) + \\
&+ \left(d_{ik} - d_{jk}\right)\left(f_{\pi_i\pi_k} - f_{\pi_j\pi_k}\right) + \left(d_{ki} - d_{kj}\right)\left(f_{\pi_k\pi_i} - f_{\pi_k\pi_j}\right) + \\
&+ \left(d_{ii} - d_{jj}\right)\left(f_{\pi_i\pi_i} - f_{\pi_j\pi_j}\right) + \left(d_{ij} - d_{ji}\right)\left(f_{\pi_i\pi_j} - f_{\pi_j\pi_i}\right) + \\
&+ \left(d_{ij} - d_{kj}\right)\left(f_{\pi_k\pi_j} - f_{\pi_i\pi_j}\right) + \left(d_{ji} - d_{jk}\right)\left(f_{\pi_j\pi_k} - f_{\pi_j\pi_i}\right) + \\
&+ \left(d_{ii} - d_{kk}\right)\left(f_{\pi_k\pi_k} - f_{\pi_i\pi_i}\right) + \left(d_{ik} - d_{ki}\right)\left(f_{\pi_k\pi_i} - f_{\pi_i\pi_k}\right) + \\
&+ \left(d_{ji} - d_{ki}\right)\left(f_{\pi_k\pi_i} - f_{\pi_j\pi_i}\right) + \left(d_{ij} - d_{ik}\right)\left(f_{\pi_i\pi_k} - f_{\pi_i\pi_j}\right) + \\
&+ \left(d_{jj} - d_{kk}\right)\left(f_{\pi_k\pi_k} - f_{\pi_j\pi_j}\right) + \left(d_{jk} - d_{kj}\right)\left(f_{\pi_k\pi_j} - f_{\pi_j\pi_k}\right)
\end{aligned}
$$

## Appendix B. MATLAB Simplification Script

```
clc
clear all
echo off

syms fii fjj fkk fij fji fik fki fjk fkj
syms dii djj dkk dij dji dik dki djk dkj


Rij = ...
    (dik - djk) * (fik - fjk) + ... % Missed g = k in delta ij
    (dki - dkj) * (fki - fkj) + ... % Missed g = k in delta ij
    (dii - djj) * (fii - fjj) + ... % Loopback
    (dij - dji) * (fij - fji);      % Reverse flows direction

Rik = ...
    (dij - dkj) * (fki - fji) + ... % Missed g = j in delta ik
    (dji - djk) * (fik - fij) + ... % Missed g = j in delta ik
    (dii - dkk) * (fkk - fjj) + ... % Loopback
    (dki - dik) * (fjk - fkj);      % Reverse flows direction

Rjk = ...
    (dki - dji) * (fij - fkj) + ... % Missed g = i in delta jk
    (dik - dij) * (fji - fjk) + ... % Missed g = i in delta jk
    (djj - dkk) * (fkk - fii) + ... % Loopback
    (dkj - djk) * (fik - fki);      % Reverse flows direction

R_ik = ...
    (dij - dkj) * (fkj - fij) + ... % Missed g = j in delta* ik
    (dji - djk) * (fjk - fji) + ... % Missed g = j in delta* ik
    (dii - dkk) * (fkk - fii) + ... % Loopback
    (dik - dki) * (fki - fik);      % Reverse flows direction

R_jk = ...
    (dji - dki) * (fki - fji) + ... % Missed g = i in delta* jk
    (dij - dik) * (fik - fij) + ... % Missed g = i in delta* jk
    (djj - dkk) * (fkk - fjj) + ... % Loopback
    (djk - dkj) * (fkj - fjk);      % Reverse flows direction


x = - Rik - Rjk + Rij + R_ik + R_jk;

simplify(x)
```